\documentclass[aps,preprint]{revtex4}
\usepackage{amsfonts}
\usepackage{amsmath}
\usepackage{amssymb}
\usepackage{graphicx}
\usepackage{subfigure}
\usepackage{mathrsfs, mathtools}
\usepackage{pdfpages}
\usepackage{multirow}
\usepackage{float}
\usepackage{color}
\usepackage{cancel}

\usepackage{xcolor}
\usepackage{orcidlink}
\usepackage{caption}
\usepackage{lineno,hyperref}
\hypersetup{colorlinks =true, allcolors = blue}
\providecommand{\U}[1]{\protect\rule{.1in}{.1in}}

\begin{document}
\begin{center}
{\Large Quasinormal modes and shadow of Schwarzschild black holes embedded in a Dehnen type dark
matter halo exhibiting string cloud}

Ahmad Al-Badawi \orcidlink{0000-0002-3127-3453}\\ 
Department of Physics, Al-Hussein Bin Talal University, P. O. Box: 20, 71111,
Ma'an, Jordan. 
\bigskip E-mail: ahmadbadawi@ahu.edu.jo\\
Sanjar Shaymatov

Institute for Theoretical Physics and Cosmology, Zhejiang University of Technology, Hangzhou 310023, China\\
Institute of Fundamental and Applied Research, National Research University TIIAME, Kori Niyoziy 39, Tashkent 100000, Uzbekistan\\
University of Tashkent for Applied Sciences, Str. Gavhar 1, Tashkent 100149, Uzbekistan\\
Western Caspian University, Baku AZ1001, Azerbaijan\\
\bigskip E-mail: sanjar@astrin.uz\\

{\Large Abstract} \end{center}
In this paper, we consider a static spherically symmetric black hole (BH) embedded in a Dehnen-(1,4,0) type dark matter (DM) halo in the presence of a cloud string. We examine and present data on how the core density of the DM halo parameter and the cloud string parameter affect BH attributes such as quasinormal modes (QNMs) and shadow cast. To do this, we first look into the effective potential of perturbation equations for three types of perturbation fields with different spins: massless scalar field, electromagnetic field, and gravitational field. Then, using the 6th order WKB approximation, we examine quasinormal modes of the BH disturbed by the three fields and derive quasinormal frequencies. The changes of QNM versus the core density parameter and the cloud string parameter for three disturbances are explored. We also investigate how the core density and the cloud string parameters affect the photon sphere and shadow radius. Interestingly, the study shows that the influence of Dehnen type DM and cloud string increases both photon spheres and shadow radius. Finally, we employ observational data from Sgr $A{^\star}$ and $M87{^\star}$ to set limitations on the BH parameters.  

\section{Introduction}

Even though Einstein’s General Relativity (GR) is a well-justified and tested theory, it is unable to provide potential explanations and insights into the inevitable occurrence of singularities inside BHs, DM (which has not been directly detected) and the accelerated expansion of the universe. From this perspective, GR has been considered an incomplete theory. However, the existence of BHs in the universe, predicted of GR, has been proven by recent observations, such as gravitational waves (GWs) \cite{Abbott16a,Abbott16b} and the BH shadow observed through the Event Horizon Telescope (EHT) \cite{Akiyama19L1,Akiyama19L6}. These observations play a pivotal role in testing the unique aspects and features of spacetime geometry in the strong field regime. It is worth noting, however, that the promising alternative theories have been proposed to provide a fundamental understanding of the issues GR faces. 

For example, the existence of DM was proposed by initially realizing the flat rotation curves of giant elliptical and spiral galaxies. It is widely believed, based on astrophysical data, that the rotational velocity of stars around giant spiral galaxies can only be explained with the help of elusive DM. It is also worth noting that DM can contribute to approximately up to 90 \% of the mass of the galaxy, while the rest is the luminous matter made op of baryonic matter \cite{Persic96}. It is also believed that, in the initial evolution of the early universe, DM was found in regions close to the center of galaxies for the formation of stars. However, in the late stages of galaxy evolution, the DM has gradually shifted out to form a DM galactic halo around the host galaxy through different dynamical scenarios. Recent astrophysical observations have reported that almost all giant elliptical and spiral galaxies have a supermassive black hole located at the galactic center with a giant DM halo \cite{Akiyama19L1,Akiyama19L6}. With this perspective, there have been various approaches to black hole solutions involving a DM distribution in the background geometry. For example, Kiselev proposed a static and spherically symmetric black hole solution with a DM profile through a quintessential scalar field \cite{Kiselev03}. Later on, Li and Yang \cite{Li-Yang12} approached this issue in a different way and eventually obtained a similar black hole solution with a DM profile represented through a phantom scalar field that contain a term $(r_q/r)\ln(r/r_q),$ with the boundary $r=r_q$ of the DM halo. It is worth noting, however, that the phantom scalar field in this model is not considered in a cosmological role. Instead, it can be modified by involving electric charge and cosmological constant \cite{Xu16-dm}, as well as spin \cite{Hou18-dm}. Following this model, an extensive analysis has since been done in a variety of contexts in the literature \cite{Hendi20,Haroon19,Rizwan19,Narzilloev20b,Shaymatov21d,RayimbaevShaymatov21a,Shaymatov21pdu}. The spherically symmetric black hole solution embedded in a DM halo can also be modeled using a Dehnen-(1, 4, 0) type DM halo \cite{dn14}. It is worth noting that this phenomenological Dehnen density profile of DM has been widely examined for galaxies (see, e.g., \cite{Dehnen93}). In this paper, we consider a Schwarzschild BH embedded in a Dehnen-(1, 4, 0) type DM halo with a string cloud background and examine its unique aspects through the astrophysical processes occurring around the BH. 

In an astrophysical scenario, it is increasingly important to explain the unique aspects and nature of the surrounding fields, such as DM, that can influence even geodesics of the massless particles, causing observable properties like the BH shadow to be affected. The key point to note is that obtaining a realistic image of the BH was a fundamental question, regardless of the fact that it had been widely investigated in the past decades. It is worth noting, however, that there has been much progress in observational studies, such as the recent detection of the first image of BH at the center of the M87*  galaxy (see, etc., \cite{Akiyama19L1,Akiyama19L6}), addressing the existence of BHs in the universe. Black holes have since taken center stage and have been tested by various theories of gravity. In this regard, the BH shadow plays an important role for observers in examining the geometry, especially very close to the BH horizon. This results in a dark disc because the light cannot escape from the pull of the BH due to the strong deflection. It must be noted that a dark disc was regarded as a BH shadow by Synge \cite{Synge66} and Luminet \cite{Luminet79}, addressing to the study of light deflection around a spherically symmetric BH. Since then, BH shadows as dark disks have since been theoretically modeled and widely examined over the past years     \cite{Amarilla13,Konoplya19,Vagnozzi19,kumar2020,Afrin21a,zhang,Atamurotov16EPJC,Konoplya19PRD,Atamurotov21JCAP,Mustafa22CPC,Tsukamoto18, Eslam-Panah20, Asukula24}. Recently, the thin accretion disk with relativistic fluid spheres was also tested extensively using observational studies \cite{Rosa23b}. We note that the unique aspects and nature of modified gravity (MOG) theory have also been tested using the BH shadows in the context of both rotating and non-rotating BH cases \cite{Moffat15EPJC,Moffat20,Al-BadawiCTP24}. The optical properties of charged black holes have also been considered within the Einstein-Maxwell-scalar (EMS) theory \cite{Al-BadawiCPC24}. Additionally, the BH shadows were also considered to obtain restrictions on theoretical models using the recent EHT observations \cite{Hendi23}.

This paper investigates a Schwarzschild BH embedded in a Dehnen-(1, 4, 0) type DM halo \cite{dn14} in the background of a cloud string. We shall look into the shadow of this BH since particle motion around BHs encodes information about not just the inherent space-time geometry but also the surrounding matter field. This necessitates studying the BH shadow in order to understand the effects of DM and string cloud. The BH shadow is essentially a black region in the celestial plane surrounded by a light emission ring. Because BH shadows are dependent on BH parameters, they are an interesting area of study. Furthermore, we will calculate QNMs, a type of gravitational wave produced by BH mergers during the ringdown phase \cite{dnn2}, with a complex characteristic frequency known as "quasinormal frequencies". The real components of the QNFs represent the perturbation's oscillation frequencies, while the imaginary parts correspond to the decay time. When a BH is disrupted, the relaxation can be represented as a superposition of exponentially damped sinusoidal signals known as QNMs \cite{dnn3}. Thus, during the ringdown stage of the coalescence of two astrophysical BHs, the GWs take the form of superposed QNMs from the remaining BH. According to the no-hair theorem, the frequencies and decay rates of these QNMs are dictated only by the physical properties of the final BH. The measurement of QNMs from GW observations would allow us to test general relativity and investigate the nature of remains from compact binary mergers.

The paper is structured as follows: Sec.~\ref{sec2} provides a concise description of the Schwarzschild BH embedded in a Dehnen Type DM Halo in the background of cloud string. Sec.~\ref{sec3} exams the QNMs of the BH disturbed by the fields with different spins and derives quasinormal frequencies. In Sec.~\ref{sec4}, we investigate the
BH shadow and we study the impact of the core density of the DM halo parameter and the cloud string parameter 
on the photon sphere and shadow radius.  Finally, our conclusions are presented in Sec.~\ref{sec6}.

\section{Review of the BH metric geometry} \label{sec2}
\begin{figure}
    \centering
    \includegraphics[width=0.5\linewidth]{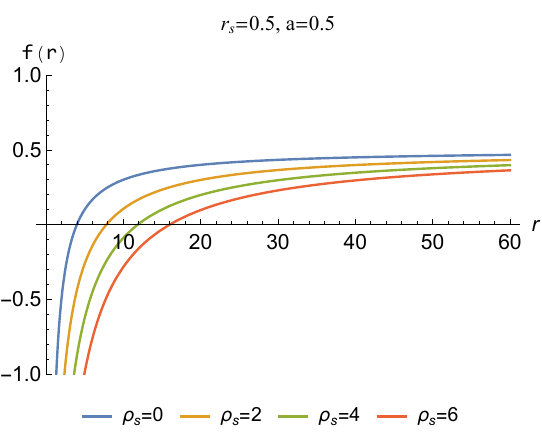}
    \caption{\label{flapse1} The function $f(r)$  is plotted for different values of the DM central density $\rho_s$ by setting the core radius and the cloud string parameter fixed.}
    \end{figure}

In this study, we will investigate the formation of a static BH with a Dehnen-type DM halo in the presence of cloud string. To do this, first the density profile of the Dehnen DM halo is a specific example of a double power-law profile defined by \cite{d1}: \begin{equation}
\rho =\rho _{s}\left( \frac{r}{r_{s}}\right) ^{-\gamma }\left[ \left( 
\frac{r}{r_{s}}\right) ^{\alpha }+1\right] ^{\frac{\gamma -\beta }{\alpha }},
\label{dens1}
\end{equation}%
here, $\rho _{s}$  is the central halo density, $r_{s}$ is the halo core
radius and $\gamma$ determines the specific variant of the profile. The values of $\gamma$ lies within
$[0, 3]$. For example, $\gamma = 3/2$ is used to fit the surface brightness profiles of elliptical galaxies which closely
resembles the de Vaucouleurs $r^{1/4}$ profile \cite{ds1}. In this paper, we use the parameters $\left( \alpha
,\beta ,\gamma \right) =\left( 1,4,0\right)$ \cite{dn14}. Therefore the above equation
becomes 
\begin{equation}
\rho _{D}=\frac{\rho _{s}}{\left( \frac{r}{r_{s}}+1\right) ^{4}}.
\label{dens2}
\end{equation}
We can obtain the mass distribution at any radial distance as 
\begin{equation}
M_{D}=4\pi \int\limits_{0}^{r}\rho \left( r^{\prime }\right) r^{\prime
2}\mathrm{d}r^{\prime }=\frac{4\pi \rho _{s}r_{s}^{3}r^3}{3\left( r+r_{s}\right) ^{3}}.
\end{equation}
 It is possible to calculate the tangential velocity of a test particle moving in the DM halo from the mass distribution of the halo profile in spherically symmetric spacetime. Thus
\begin{equation}
v_{D}^{2}=\frac{M_{D}}{r}=\frac{4\pi \rho _{s}r_{s}^{3}r^2}{3r\left( r+r_{s}\right) ^{3}}.
\end{equation}
Assume the following for a static spherically symmetric line element describing a pure DM halo:
\begin{equation}
\mathrm{d}s^{2}=-F\left( r\right) \mathrm{d}t^{2}+\frac{\mathrm{d}r^{2}}{F\left( r\right) }+r^{2}\left(
\mathrm{d}\theta ^{2}+\sin ^{2}\theta \mathrm{d}\phi ^{2}\right).\label{m2}
\end{equation}%
In Eq. (\ref{m2}) the metric function $F(r)$ is related to the tangential velocity through the
expression%
\begin{equation}
v_{D}^{2}=r\frac{\mathrm{d}}{\mathrm{d}r}\left( \ln \sqrt{F(r)}\right) .
\end{equation}%
Solving for $F(r)$ we obtain 
\begin{equation}
F(r)=exp\left[- \frac{4\pi \rho _{s}r_{s}^{3}\left( 2r+r_{s}\right) }{%
3\left( r+r_{s}\right) ^{2}}\right] \approx 1-\frac{4\pi \rho
_{s}r_{s}^{3}\left( 2r+r_{s}\right) }{3\left( r+r_{s}\right) ^{2}},
\end{equation}
where the leading order terms of the equation were retained. The next step is to combine the DM profile encoded in $F(r)$ with the BH metric function. We used Xu et al.'s formalism \cite{dn9}, which has also been used by others \cite{dn10,dn11,dn12,dn13,dn14}, to write the metric line element in this scenario. Thus, the metric of a static and spherically symmetric BH embedded in DM halo exhibiting string cloud is described by a line element \cite{dn14,ds4} 
\begin{equation} 
\mathrm{d}s^{2}=-f\left( r\right) \mathrm{d}t^{2}+\frac{\mathrm{d}r^{2}}{f\left( r\right) }+r^{2}\left(
\mathrm{d}\theta ^{2}+\sin ^{2}\theta \mathrm{d}\phi ^{2}\right) ,\label{m1}
\end{equation} where 
\begin{equation}
f\left( r\right) =1-a-\frac{2M}{r}-\frac{4\pi \rho _{s}r_{s}^{3}\left(
2r+r_s\right) }{3\left( r+r_{s}\right) ^{2}},  \label{laps1}
\end{equation}
where $a$ refers to the string cloud parameter. 
Figure \ref{flapse1} shows a visualisation of the Lapse function (\ref{laps1}) as a function of $r$. The plot shows that the centre density of the DM halo has a major influence on the occurrence of BH horizons while the core radius and the cloud string parameter remain constants at $r_s = 0.5=a$. The horizons can be found using the constraint $f(r) = 0$. Accordingly, there exists a unique horizon for the combinations of dark matter halo and the cloud string.

\section{quasinormal modes} \label{sec3}

\begin{figure}[H]
\begin{center}
\includegraphics[scale=0.5]{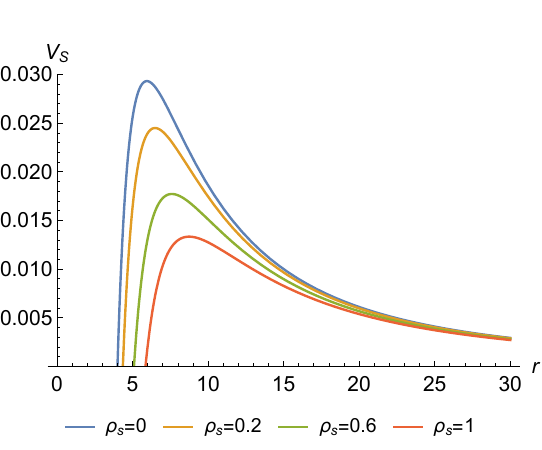}
\includegraphics[scale=0.5]{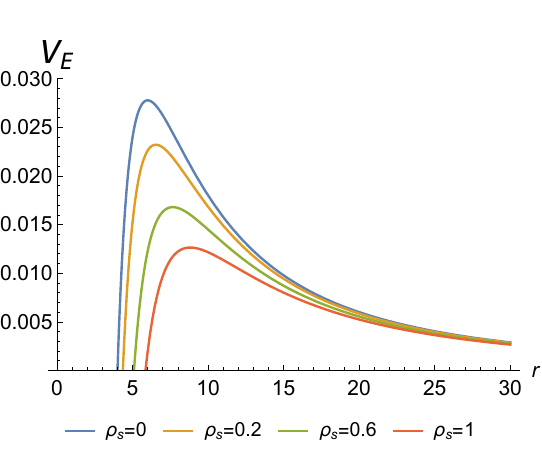}
\includegraphics[scale=0.5]{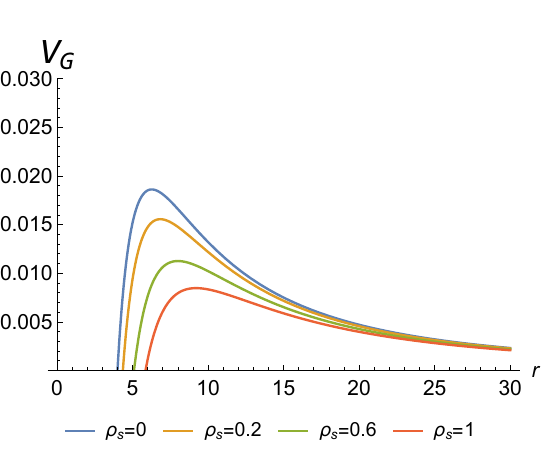}
\end{center}
\caption{Behaviours of BH potential with respect to radial distance $r$ for different values of central density of the DM halo $\rho_s$. Here, $M=1,l=2$ and $r_s=0.5=a$. }\label{fpotRo}
\end{figure}

\begin{figure}[H]
\begin{center}
\includegraphics[scale=0.5]{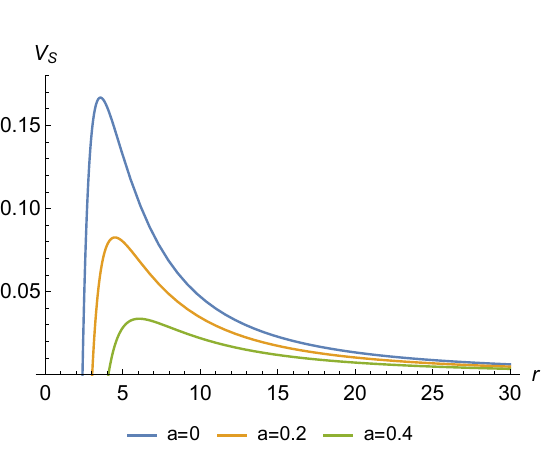}
\includegraphics[scale=0.5]{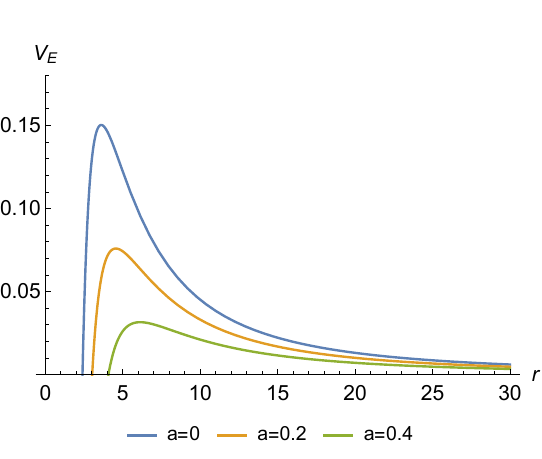}
\includegraphics[scale=0.5]{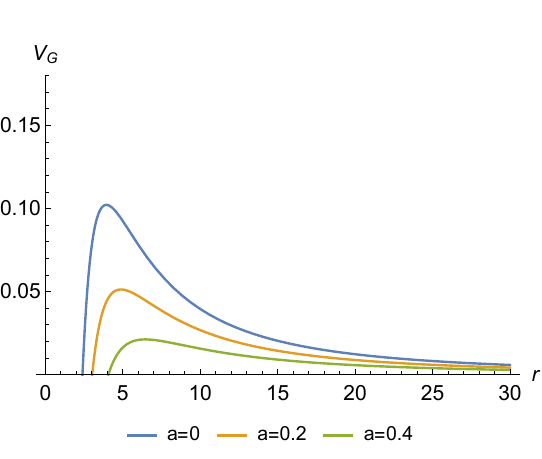}
\end{center}
\caption{Behaviours of BH potential with respect to radial distance $r$ for different values of cloud string $a$. Here, $M=1,l=2$ and $r_s=0.5=\rho_s$. }\label{fpotA}
\end{figure}

In this part, we compute the QNMs belonging to the Schwarzschild BH embedded in a Dehnen type DM halo exhibiting string cloud spacetime provided in Eq. (\ref{m1}). We use the most standard method, the 6th-order WKB approximation \cite{d3,d4,Konoplya19WKB}. We shall investigate the perturbation equations for three types of perturbation fields with varying spins: massless scalar field (S), electromagnetic field (E), and gravitational field (G).  The field equations of the scalar $\Phi$ and electromagnetic  $A_{\mu}$ field can be described as follows:  
\begin{equation}
\frac{1}{\sqrt{-g}}\partial _{\mu }\left( \sqrt{-g}g^{\mu \nu }\partial
_{\mu }\Phi \right) =0,  \label{eq21}
\end{equation}
\begin{equation}
\frac{1}{\sqrt{-g}}\partial _{\mu }\left( F_{\alpha \beta }g^{\alpha \nu
}g^{\beta \mu }\sqrt{-g}\right) =0,  \label{eq22}
\end{equation}
where $F_{\alpha \beta }=\partial _{\alpha }A_{\beta }-\partial _{\beta
}A_{\nu }$ is the electromagnetic tensor. 
The general master equation of a static symmetric BH for calculating quasinormal modes is:
\begin{equation}
\frac{\mathrm{d}^{2}\Psi}{\mathrm{d}r_{\ast }^{2}}+\left( \omega^{2}-V(r_{\ast })\right) \Psi=0,  \label{s3}
\end{equation}
We will assume the perturbations depend on
time as $e^{-i\omega t}$. Thus, to have damping, $\omega$ must have a negative imaginary part. The tortoise coordinate $r_{\ast }$ is defined by $r_{\ast } = \int \frac{dr}{f}$. The effective potentials $V(r_{\ast })$ for three types of perturbation fields in Eq. (\ref{s3}) are  
\begin{equation}
V_{S}(r)=\left(\frac{1}{2}+l\right)\frac{f }{r^{2}}+\frac{ff^{\prime }}{r}, \label{pots}
\end{equation}%
\begin{equation}
V_{E}(r)=\left(\frac{1}{2}+l\right)\frac{f }{r^{2}},  \label{pots10} 
\end{equation}
\begin{equation}
V_{G}(r)=\left(\frac{1}{2}+l\right)\frac{f }{r^{2}}-2f\frac{1-f}{r^2}- \frac{ff^{\prime }}{r} ,
\label{pot11}
\end{equation}
where $l$ is the standard spherical harmonics indices. 
Figures \ref{fpotRo} and \ref{fpotA} show graphs of the potentials (\ref{pots}-\ref{pot11}) to explore both the effect of the DM halo density as well as the cloud of string parameter. The figures show that both the core density of the DM halo and the cloud string parameter have a considerable effect on the potentials.  Potentials decrease when
$\rho_s$  and  $a$ grow. However, the cloud string parameter has more influence than the DM halo. 
The next step is to compute the quasinormal frequencies of the Schwarzschild BH embedded in a Dehnen type DM halo with a string cloud.    To achieve physical consistency at both the BH horizon and infinity, Equation (\ref{s3}) needs to be subject to the necessary boundary conditions. To have asymptotically flat spacetime, the following quasinormal criteria must be met.  \begin{equation}
\Psi \left( r_{\ast }\right) \rightarrow \left\{ 
\begin{array}{cc}
Ae^{i\omega r_{\ast }}, & r_{\ast }\rightarrow -\infty  \\ 
Be^{-i\omega r_{\ast }}, & r_{\ast }\rightarrow \infty 
\end{array}%
\right. 
\end{equation}
The coefficients $A$ and $B$ represent the wave amplitudes. These ingoing and outgoing waves are consistent with the physical requirements that nothing can leave the BH's horizon and no radiation can arise from infinity. Furthermore, these guarantee the existence of an infinite set of discrete complex numbers, often known as QNMs. \\ We focus on the fundamental mode with $l = 2$ and overtone
number $n = 0$ due to its dominant ingredient of gravitational waves in order to study the impact of Dehnen type DM parameters and string cloud parameter on the quasinormal frequencies. As seen in tables \ref{taba1} and \ref{taba2}, the frequencies derived by 6th order WKB approximation have negative imaginary parts, demonstrating stability under various perturbations. 
\begin{center}
\begin{tabular}{|c|c|c|c|}
 \hline 
 \multicolumn{4}{|c|}{ $r_{s}=0.5$, $a=0.5$ $n=0$, $M=1$}
\\ \hline $\rho _{s}$ & $Scalar$ & EM & Gravitational \\ \hline
$0$ & $0.168842-0.02412i$ & $0.164225-0.023904i$ & $0.133632-0.022988i
$ \\ 
$0.2$ & $0.15435-0.021927i$ & $0.15016-0.021729i$ & $0.122193-0.020899i$ \\ 
$0.4$ & $0.141975-0.02008i$ & $0.13813-0.019902i$ & $0.112407-0.019144i$ \\ 
$0.6$ & $0.13130-0.018513i$ & $0.12775-0.018348i$ & $0.103965-0.017651i$ \\ 
$0.8$ & $0.12202-0.017165i$ & $0.11874-0.017013i$ & $0.096628-0.016368i$ \\ 
$1$ & $0.11390-0.015996i$ & $0.11084-0.015855i$ & $0.090204-0.015254i$%
\\ 
 \hline
\end{tabular}
\captionof{table}{Variation of amplitude and damping of QNMs with respect to central halo density parameter.} \label{taba1}
\end{center}

\begin{figure}
    \centering
    \includegraphics[width=18cm]{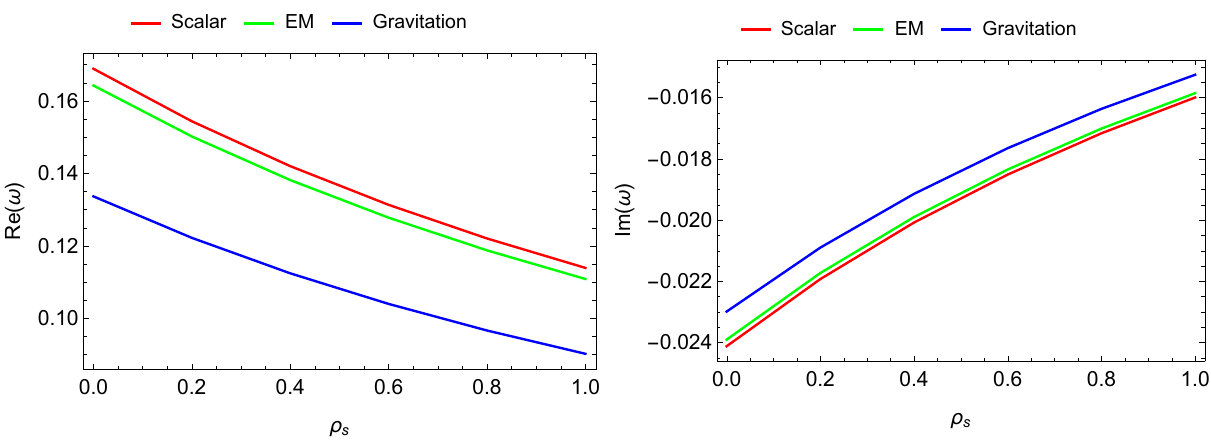}
    \caption{ Variation of amplitude and damping of QNMs with respect to the central density of the DM halo  parameter  for three different perturbations.  Here, $M=1$, $a=0.5$ and $r_s=0.5$.}
    \label{realRo}
\end{figure}

\begin{figure}
    \centering
    \includegraphics[width=18cm]{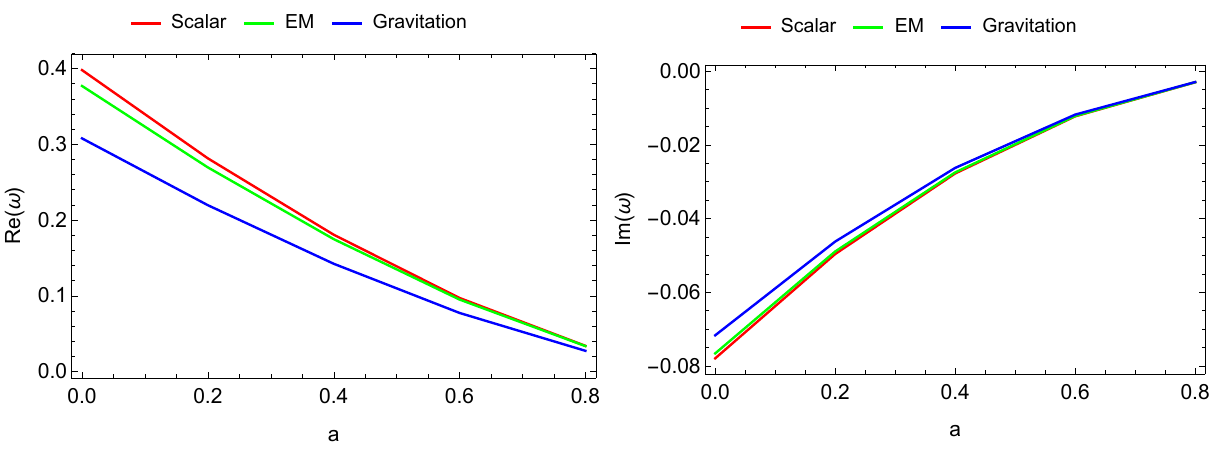}
    \caption{ Variation of amplitude and damping of QNMs with respect to the central density of the DM halo  parameter  for three different perturbations.  Here, $M=1$, $\rho=0.5$ and $r_s=0.5$.}
    \label{realA}
\end{figure}

\begin{figure}
    \centering
  \includegraphics[scale=0.85]{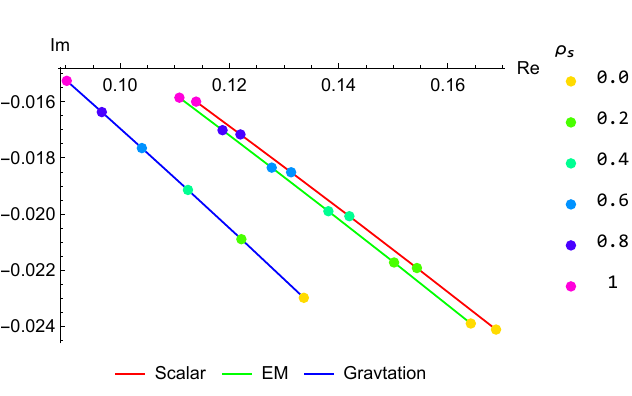}
    \caption{Complex frequency plane for the Scalar, EM and Dirac perturbations  showing the behavior of the quasinormal frequencies.  Here, $M=1$, $a=0.5$ and $r_s=0.5$.}
    \label{imagR}
\end{figure}

\begin{figure}
    \centering
  \includegraphics[scale=0.85]{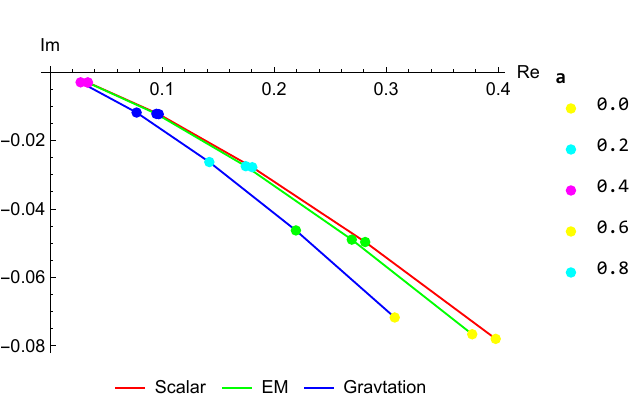}
    \caption{Complex frequency plane for the Scalar, EM and Dirac perturbations  showing the behavior of the quasinormal frequencies.  Here, $M=1$, $\rho=0.5$ and $r_s=0.5$.}
    \label{imagA}
\end{figure}

 \begin{center}
\begin{tabular}{|c|c|c|c|}
 \hline 
 \multicolumn{4}{|c|}{  $r_{s}=0.5$, $\rho _{s}=0.5$, $n=0$, $M=1$}
\\ \hline          
$a$ & $Scalar$ & EM & Gravitational \\ \hline
$0$ & $0.397707-0.077968i$ & $0.376752-0.076623i$ & $0.307594-0.071715i$ \\ 
$0.2$ & $0.281252-0.049663i$ & $0.269282-0.048969i$ & $0.219422-0.046280i$ \\ 
$0.4$ & $0.180477-0.027806i$ & $0.174658-0.027511i$ & $0.142158-0.026297i$ \\ 
$0.6$ & $0.097018-0.012303i$ & $0.094911-0.012215i$ & $0.077240-0.011830i$ \\ 
$0.8$ & $0.033862-0.003063i$ & $0.033490-0.003051i$ & $0.027282-0.003000i$%
\\ 
 \hline
\end{tabular}
\captionof{table}{Variation of amplitude and damping of QNMs with respect to cloud string parameter.} \label{taba2}
\end{center}
Figures \ref{realRo} and \ref{realA} show the relationship of QNM with the central density of the DM halo parameter and the string cloud parameter. According to the figures, QNM amplitude and damping generally increase with the parameters $\rho_s$ and $a$. However, the effect of $a$ on quasinormal frequencies is greater than $\rho_s$. Figures \ref{imagR} and \ref{imagA} illustrate the quasinormal frequencies in the complex frequency plane. The graphic shows that the general trend of amplitude and damping of QNMs is that both increase with the parameter $\rho_s$ and $a$.

\section{Shadow} \label{sec4}

\begin{figure}
\begin{center}
\includegraphics[scale=0.8]{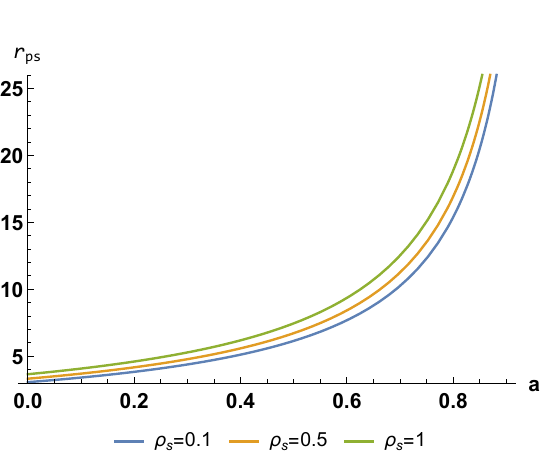}
\includegraphics[scale=0.8]{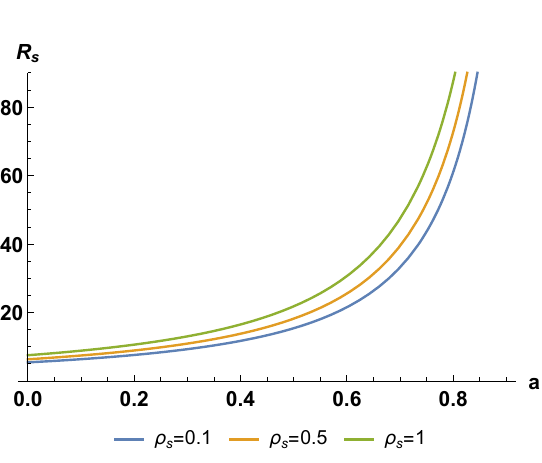}
\end{center}
\caption{Dependence of $r_{ps}$ and $R_s$ with respect to cloud string $a$ for different values of $\rho_s$. Here, $M=1$ and $r_s=0.5$. }\label{shadow32}
\end{figure}

\begin{figure}
\begin{center}
\includegraphics[scale=0.55]{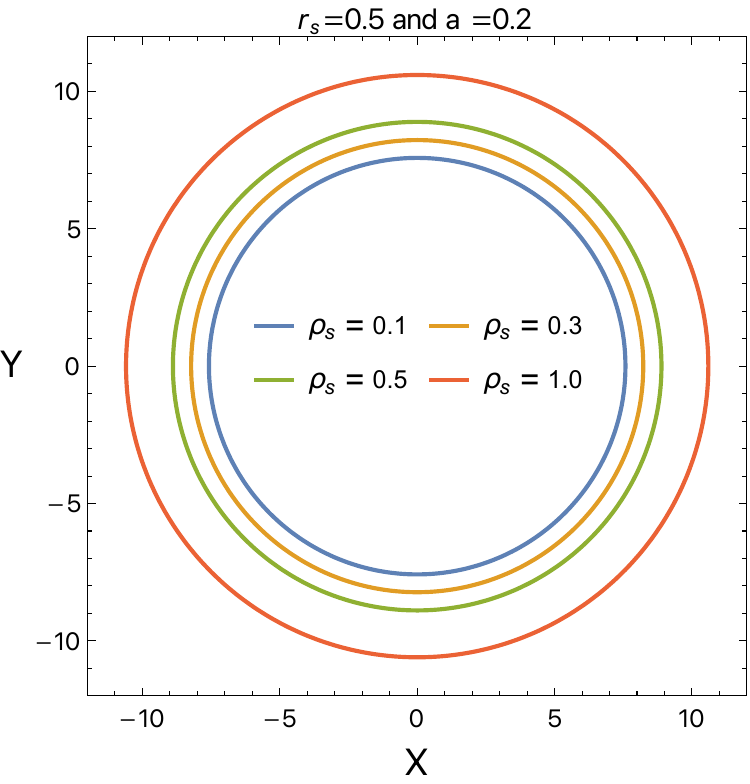}
\includegraphics[scale=0.55]{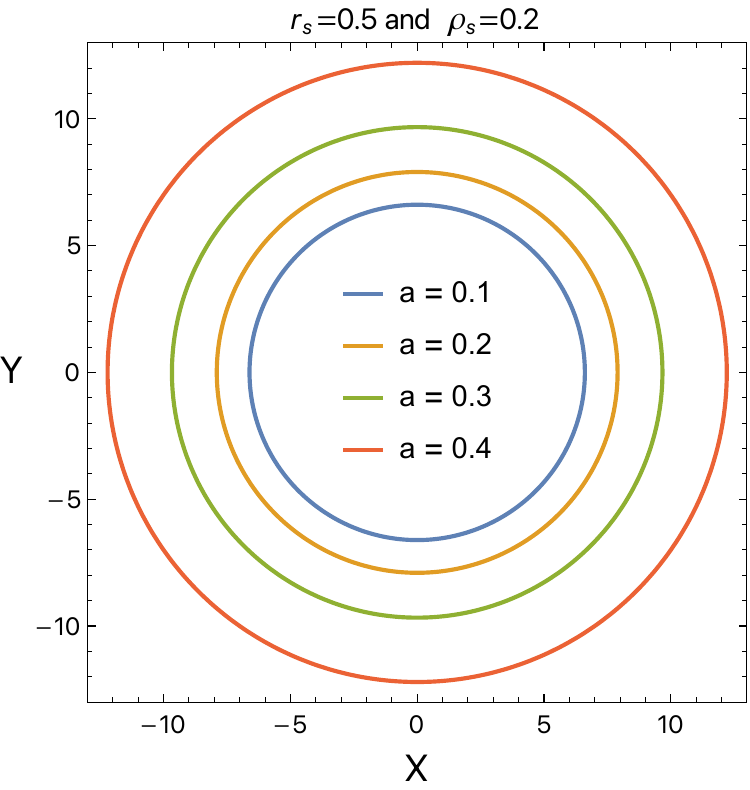}
\end{center}
\caption{Profile of shadows by the Schwarzschild BH embedded in a Dehnen type DM halo with the string of cloud for different values of $\rho_s$ (left) and $a$ (right). Here, we note that we have set $M=1$ and $r_s=0.5$. }\label{shadow33}
\end{figure}

  The BH shadow has been extensively explored in the literature because it provides a good opportunity to test various gravity and BH theories under extreme gravity regimes. By using observational data on BH shadow radius, the scientific community has been able to constrain model parameters. In this section, we compute and plot the photon sphere $r_{ps}$ and shadow radius $R_s$ expressions to analyse their reliance on the DM central density $\rho_s$ and string cloud parameter $a$. We also try to limit the parameter space using observational data from the EHT group. \\ The motion of light can be found by using the Euler–
Lagrange equation as follows \begin{equation}
    \frac{\mathrm{d}}{\mathrm{d}\tau}\!\left(\frac{\partial\mathcal{L}}{\partial\dot{x}^{\mu}}\right)-\frac{\partial\mathcal{L}}{\partial
x^{\mu}}=0, 
\end{equation}
 where the generalized Lagrangian is given by 
\begin{equation}
\mathcal{L}(x,\dot{x})=\frac{1}{2}\,g_{\mu\nu}\dot{x}^{\mu}\dot{x}^{\nu}.
\end{equation} 
 We confine ourselves to the
equatorial plane $(\theta=\pi/2)$ and define the two constant of motion as 
 $E=f(r)\dot{t}$, and $L=r^2 \dot{\phi}$.  Thus, the
equation of the effective potential \begin{equation}
    V_{eff}=\frac{f}{r^2}\left( \frac{L^2}{E^2}-1\right).
\end{equation} 
  The radius of an unstable photon orbit can be obtained using the following conditions $V_{eff}=0=V'_{eff}$  though 
the photon sphere radius of a BH is calculated from \begin{equation}
    r_{ps}f'(r_{ps})-2f(r_{ps})=0. 
\end{equation}
Therefore, the equation of BH's photon sphere radius is given by  \begin{equation}
   \big( 9M-3r_{ps}(1-a)\big)(r_{ps}+r_s)^3+4\pi \rho_{s}r_{s}^{3}r_{ps}\left(
3r_{ps}^2+3r_{ps}r_s+r_{s}^2\right)=0.\label{ps1}
\end{equation}
 As a result of the photon radius, we can derive the critical impact parameter or the shadow radius $R_s$ as follows: \begin{equation}
     R_s=\frac{L}{E}=\frac{r_{ps}}{\sqrt{f(r_{ps})}}.\label{sr1}
 \end{equation} 
Equations (\ref{ps1}, \ref{sr1}) show how the parameters $\rho _{s}$ and $a$ affect photon motion in the Schwarzschild BH embedded in a Dehnen type DM
 halo exhibiting string cloud. In either case, it is not possible to obtain analytical expressions for the photon radius or the shadow radius. Therefore, we compute numerical values for photon radius and associated shadow radius in order to have a better grasp of the impact. The Table (\ref{taba3}) clearly shows that both radii increase with parameters $\rho _{s}$ and $a$.
 \begin{center}
\begin{tabular}{|c|c|c|c|c|c|c|}
 \hline 
 &\multicolumn{2}{|c|}{  $\rho _{s}=0.1$}& \multicolumn{2}{|c|}{  $\rho _{s}=0.5$}& \multicolumn{2}{|c|}{  $\rho _{s}=1$}
\\ \hline
$a$ & $r_{ps}$ & $R_{s}$ & $r_{ps}$ & $R_{s}$ & $r_{ps}$ & $R_{s}$ \\ \hline
$0$ & $3.11752$ & $5.41447$ & $3.6093$ & $6.31938$ & $4.26351$ & $7.50642$
\\ 
$0.1$ & $3.46745$ & $6.34681$ & $4.02703$ & $7.42592$ & $4.76781$ & $8.83591$
\\ 
$0.2$ & $3.90505$ & $7.57985$ & $4.54985$ & $8.89104$ & $5.39906$ & $10.5972$
\\ 
$0.3$ & $4.4679$ & $9.26911$ & $5.22283$ & $10.9006$ & $6.21166$ & $13.0143$
\\ 
$0.4$ & $5.21863$ & $11.6913$ & $6.12106$ & $13.7852$ & $7.29632$ & $16.4859$
\\ 
$0.5$ & $6.27$ & $15.3834$ & $7.37975$ & $18.1871$ & $8.81628$ & $21.7863$
\\ 
$0.6$ & $7.84751$ & $21.5204$ & $9.26928$ & $25.512$ & $11.098$ & $30.611$
\\ 
$0.7$ & $10.4773$ & $33.1672$ & $12.4205$ & $39.4281$ & $14.9034$ & $47.385$
\\ 
$0.8$ & $15.738$ & $60.9976$ & $18.7263$ & $72.7165$ & $22.5177$ & $87.5309$%
\\ 
 \hline
\end{tabular}
\captionof{table}{Numerical values of $r_{ps}$ and $R_s$.  Here, $M=1$ and $r_s=0.5$.} \label{taba3}
\end{center}
 Figure \ref{shadow32}  illustrates how $r_{ps}$ and $R_s$ are influenced by DM core density and string cloud parameter. The figure illustrates how $\rho_s$ and $a$ have similar effects on the photon sphere and shadow radius. The existence of Dehnen type DM and string cloud causes the both $r_{ps}$ and $R_s$ to increase.  \\ Now we consider the geometrical quantity on a celestial plane along the coordinates $X$ and $Y$ given by \cite{Vazquez}:
\begin{equation}
X=\lim_{r_{0}\rightarrow \infty }\left( -r_{0}\sin \theta _{0}\left. \frac{%
\mathrm{d}\phi }{\mathrm{d}r}\right\vert _{r_{0},\theta _{0}}\right) ,  \label{x11}
\end{equation}%
\begin{equation}
Y=\lim_{r_{0}\rightarrow \infty }\left( r_{0}\left. \frac{\mathrm{d}\theta }{\mathrm{d}r}%
\right\vert _{r_{0},\theta _{0}}\right) ,  \label{y11}
\end{equation}%
where $r_0$ denotes the distance between the BH and the observer. In two-dimensional geometry governed by the shadow radius, Eq. (\ref{y11}) can be expressed as follows:
\begin{equation}
X^{2}+Y^{2}=R_{s}^{2}=\eta +\zeta ^{2}.  \label{xy1}
\end{equation}
We show how the shadow size varies with core density and string cloud parameter in figure \ref{shadow33}. 
  
\begin{figure}
    \centering
    \includegraphics[scale=0.55]{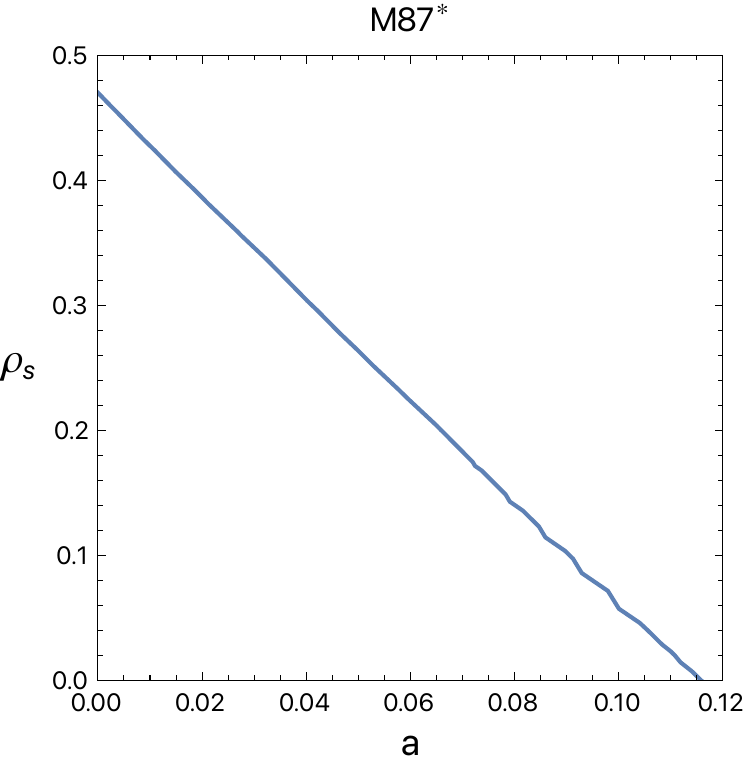}%
    \includegraphics[scale=0.55]{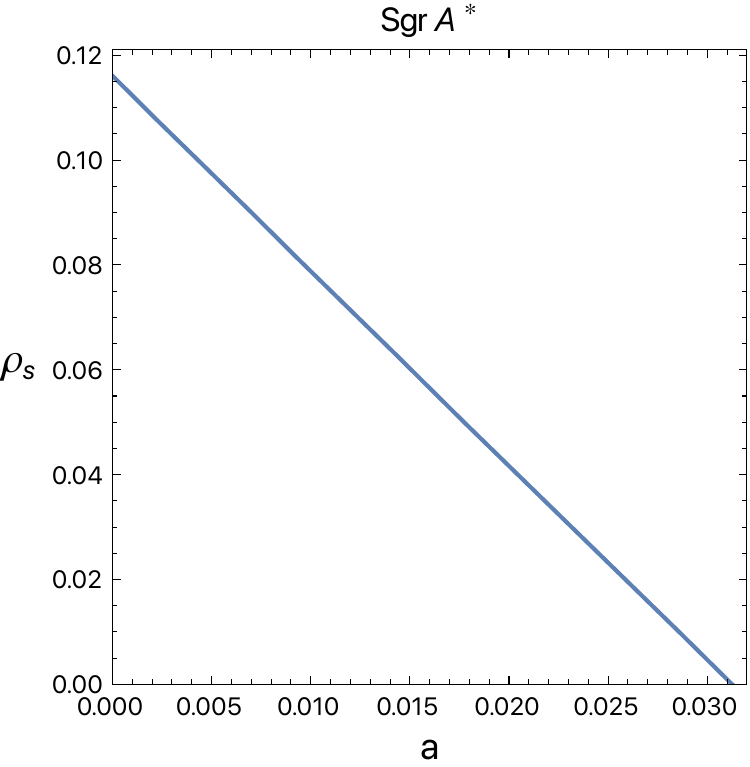}
    \caption{Constraint values of the DM density $\rho_s$ and string cloud $a$ for M87$^{\star}$ and Sgr A$^{\star}$. Here, we note that we have set $M=1$ and $r_s=0.5$. }
    \label{fig:constraint}
\end{figure}
\begin{figure}
    \centering
    \includegraphics[width=0.45\textwidth]{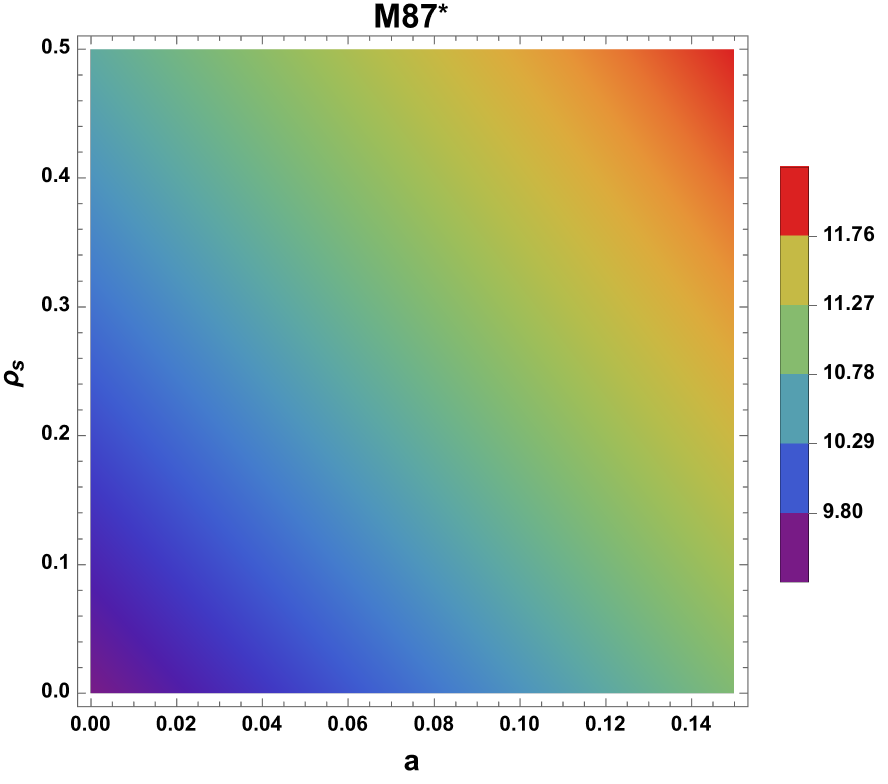}%
    \includegraphics[width=0.465\textwidth]{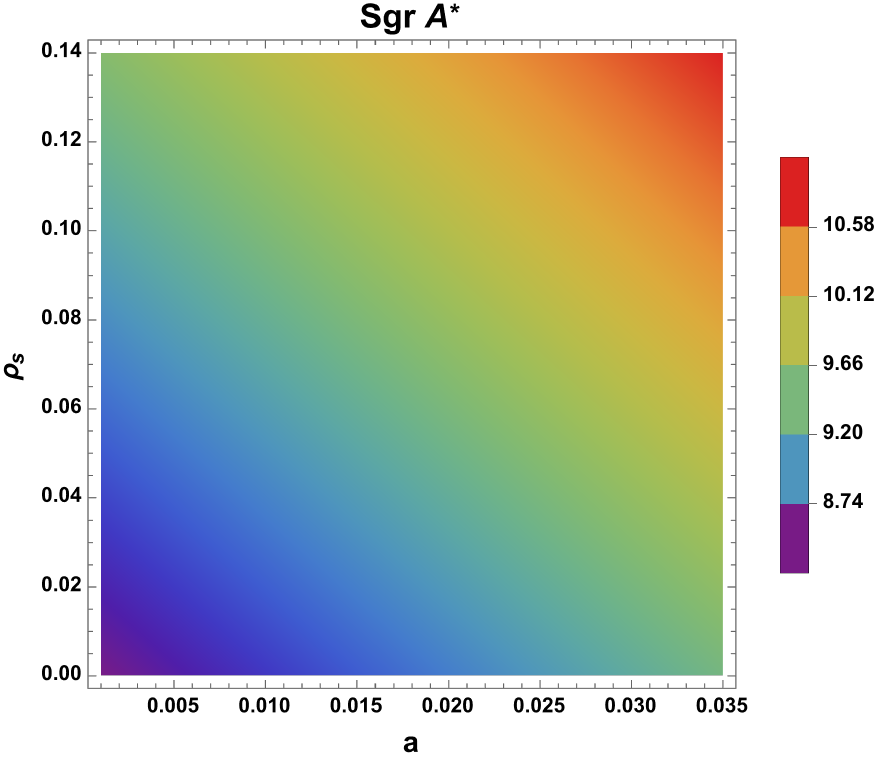}
    \caption{The density plot of the DM parameter $\rho_s$ and string cloud $a$ using the data of M87$^{\star}$ and Sgr A$^{\star}$. Here, we note that we have set $M=1$ and $r_s=0.5$. }
    \label{fig:density}
\end{figure}
\begin{figure}
    \centering
    \includegraphics[scale=0.55]{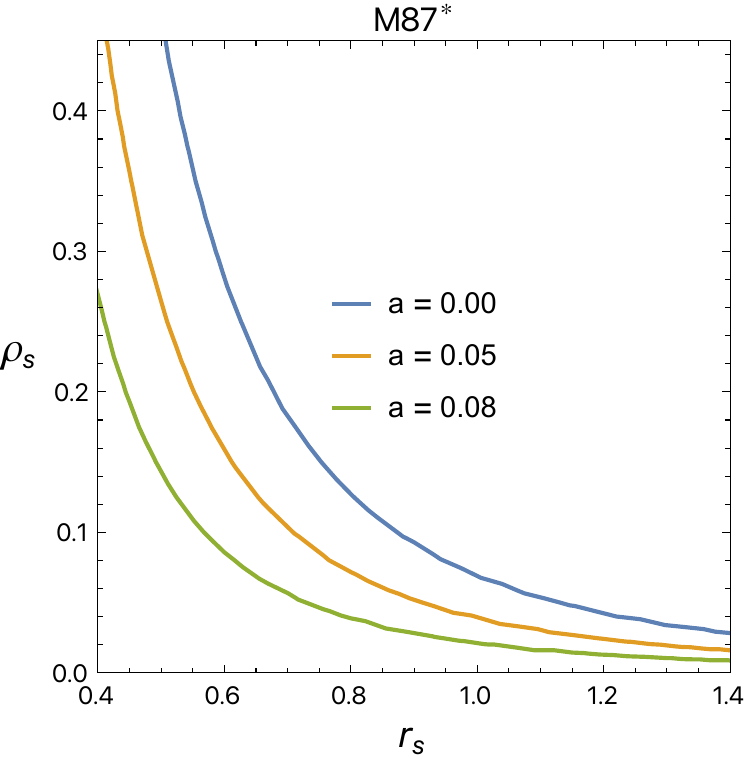}%
    \includegraphics[scale=0.55]{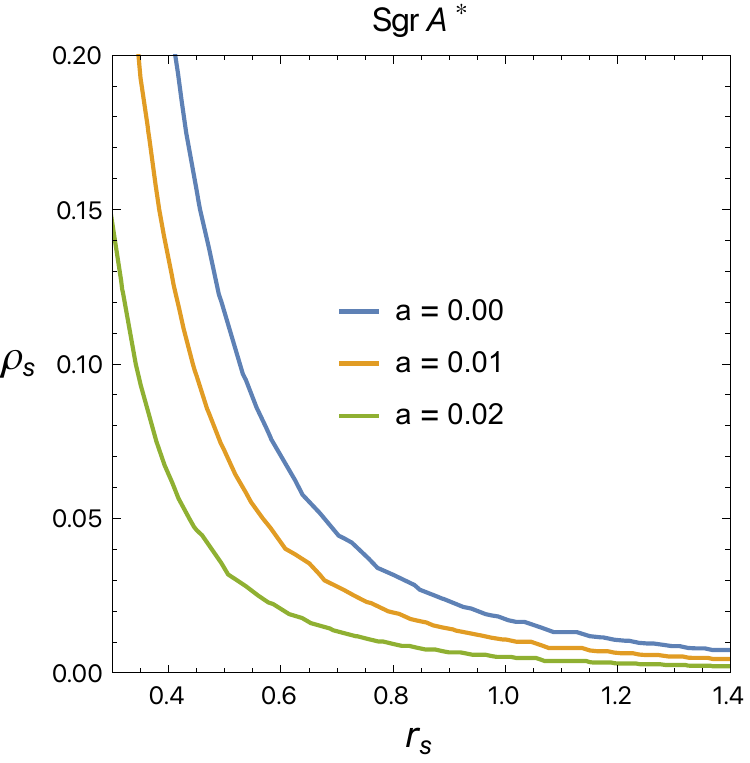}
    \caption{Constraint values of the DM density $\rho_s$ and the halo core radius $r_s$ for M87$^{\star}$ and Sgr A$^{\star}$ for various combinations of string cloud parameter $a$. Here, we note that we have set $M=1$.} 
    \label{fig:constraint2}
\end{figure}

We will now explore the astrophysical implications of the obtained theoretical results using the recent EHT observations, which provide upper limits for the BH parameters. With this in mind, we can assume that supermassive BHs such as Sgr A$^{\star}$ and M87$^{\star}$ are static and spherically symmetric BHs in our theoretical model considered here. Additionally, we assume that they support our approach and the assumptions we have made here. We then turn to constrain the lower limits of the parameters $\rho_s$ and $a$ as stated by the EHT observational data associated with the BH shadow's angular diameter $\theta$, the distance $D$, and the mass of supermassive BHs located at the center of Sgr A$^{\star}$ and M87$^{\star}$ galaxies. Based on the observational data, these parameters are given for M87$^{\star}$ and Sgr A$^{\star}$ as follows: $\theta_{M87^{\star}}=42 \pm 3 \mu as$, $D= 16.8 \pm 0.8 M pc$ between Earth and M87$^{\star}$ with mass $M_{M87^{\star}} = (6.5 \pm 0.7) \times 10^9 M_{\odot}$ and $\theta_{Sgr A^{\star}}=48.7 \pm 7 \mu$,  $D=8277 \pm 9 \pm 33 pc$ and $M_{Sgr A^{\star}} = (4.297 \pm 0.013) \times 10^6 M_{\odot}$ (see details in \cite{Akiyama19L1,Akiyama19L6}). Taken altogether, we estimate the diameter of the BH shadow per unit mass, relying on the following equation
\begin{eqnarray}
    d_{sh}=\frac{D\,\theta}{M}\, .
\end{eqnarray}
Taking $d_{sh}=2R_{sh}$ into further consideration, we determine the diameter of the BH shadow by defining $d^{M87^{\star}}_{sh}=(11 \pm 1.5)M$ for M87$^{\star}$ and $d^{Sgr^{\star}}_{sh}=(9.5 \pm 1.4)M$ for Sgr A$^{\star}$, respectively. Following the observational EHT data, we examine the parameters $\rho_s$ and $a$ for both Sgr A$^{\star}$ and M87$^{\star}$ cases and demonstrate their upper values in Fig.~\ref{fig:constraint}. It is clearly seen from Fig.~\ref{fig:constraint} that the upper values of $\rho_s$ and $a$ from observational data of Sgr A$^{\star}$ are smaller compared to those from M87$^{\star}$. We exhibit the comparison between the data of M87$^{\star}$ and Sgr A$^{\star}$. Therefore, we can deduce that the upper value of the DM parameter $\rho_s$ decreases as the string cloud parameter $a$ increases. However, the effect of both parameters would be much more prominent on the background geometry. We also show the density plot of the parameters $\rho_s$ and $a$ by applying the data of M87$^{\star}$ and Sgr A$^{\star}$ in Fig.~\ref{fig:density}. Here, we can observe that the behavior of the parameters $\rho_s$ and $a$ is also summarized in Fig.~\ref{fig:density}. From observational data of M87$^{\star}$ and Sgr A$^{\star}$, one can estimate the approximate range of the parameters $\rho_s$, $r_s$, and $a$. We demonstrate the range of parameters $\rho_s$ and $r_s$ for various possible cases of the string cloud parameter $a$ in Fig.~\ref{fig:constraint2}. As shown in Fig.~\ref{fig:constraint2}, the dark matter density $\rho_s$ is well distributed in the range of $0.5\leq r_s \leq1$ around the BH. It should be emphasized that the \cancel{\textcolor{red}{minimum}} range of $r_s$ can be extended up to its possible \textcolor{red}{small values} \cancel{\textcolor{red}{minimum}}, but the DM density profile $\rho_s$ decreases when considering the string cloud $a$, as depicted in Fig.~\ref{fig:constraint2}. It is worth noting, however, that the DM density parameter $\rho_s$ can take values up to $<0.5$ and $<0.12$ for M87$^{\star}$ and Sgr A$^{\star}$, respectively. With this, we can infer from the theoretical results that these observational data of M87$^{\star}$ and Sgr A$^{\star}$ are potentially important for the best-fit constraints on the BH parameters such as $\rho_s$, $r_s$ and $a$ and can help constrain the estimations of the DM profile around astrophysical BHs.

\section{Conclusion} \label{sec6}

The combination of BH and DM in the background of a string cloud can be an exciting physical system, potentially crucial in terms of BH properties such as QNMs and shadow cast. In the first section of the study, we derived the functional form of the BH metric function by examining a Schwarzschild BH immersed in a Dehnen-type DM halo profile with a cloud string. We plotted the metric function vs $r$ for various model parameter values and find that the central density of the DM halo has a considerable influence on the presence of BH horizons while maintaining the core radius and string cloud parameter constant ($r_s=0.5=a$). It was shown that the combination of the dark matter halo and string cloud results in a unique horizon. \\We then investigated the effective potential of perturbation equations for three types of perturbation fields with varying spins: massless scalar, electromagnetic, and gravitational fields. Using the 6th order WKB approximation, we investigated quasinormal modes of the BH disrupted by the three fields and calculated quasinormal frequencies. The effects of QNM on the core density of the DM halo parameter and the cloud string parameter for three disturbances are investigated. It is found that the core density of the DM halo parameter and the cloud string have opposing effects on the QNM amplitude and damping. Calculations indicate that QNM amplitude and damping increase with $\rho_s$ and decrease with $a$, respectively. Tables \ref{taba1} and \ref{taba2} show tabulated QNM data that support this pattern. Both $\rho_s$ and $a$ clearly have an effect on the QNM frequencies. Due to the fact that gravitational waves have a higher frequency and decay rate with $\rho_s$ and $a$ for the three perturbations, gravitational waves emitting from BHs embedded in a Dehnen-(1,4,0) type DM halo in the presence of cloud string will propagate further at smaller values of $\rho_s$ and smaller values of $a$. \\
We also examined how the core density of the DM halo parameter and the cloud string parameter affected the photon sphere and shadow radius.  The study found that both the cloud string and core density parameters have similar  effects on the photon sphere and shadow radius. Both the photon sphere and the shadow radius increase as a consequence of an increase in the value of $\rho_s$ and $a$. However, the existence of cloud string has greater impact on shadow size. Since we have shown that the shadow size is dependent on BH parameters, we can utilize the obtained results to establish constraints on the parameters of a BH embedded in a Dehnen-type DM halo in the presence of cloud strings. Using observational data from M87$^{\star}$ and Sgr A$^{\star}$, we identified a possible range of constraints on the parameters $\rho_s$ and $r_s$ in the presence of the string cloud parameter $a$. Our analysis indicates that the DM density $\rho_s$ can be well distributed within the range of $0.5\leq r_s \leq1$ around the BH. However, the minimum range of $r_s$ can be narrowed down to its possible minimum when considering the string cloud parameter $a$. It is also noteworthy that the DM density $\rho_s$ can fall within the ranges of $<0.5$ and $<0.12$ with string cloud parameter $a<0.12$ and $<0.032$ for M87$^{\star}$ and Sgr A$^{\star}$, respectively. \\ Given this study, it is intriguing to explore the impact of the Dehnen profile in the presence of a string cloud on the weak and strong deflection angles of massive particles, which we would next intend to examine in a separate study.

\section*{ACKNOWLEDGEMENT}

This work is supported by the National Natural Science Foundation of China under Grant No. 11675143 and the National Key Research and Development Program of China under Grant No. 2020YFC2201503.

\end{document}